\documentclass[%
 reprint,
superscriptaddress,
 amsmath,amssymb,
 aps,
]{revtex4-2}

\usepackage{graphicx}% Include figure files
\usepackage{dcolumn}% Align table columns on decimal point
\usepackage{bm}% bold math
\DeclareMathOperator\erf{erf}
\newcommand{\NEELAddress}{Univ. Grenoble-Alpes, CNRS, Inst. NEEL, Grenoble, France}
\newcommand{\IRIGAddress}{Univ. Grenoble-Alpes, CEA, IRIG, Grenoble, France}

\begin{document}

\preprint{APS/123-QED}

\title{Quantitative analysis of the blue-green single-photon emission \\from a quantum dot in a thick tapered nanowire}% Force line breaks with \\

\author{Saransh Raj Gosain}
\author{Edith Bellet-Amalric}
\author{Eric Robin}
\affiliation{\IRIGAddress}
\author{Martien Den Hertog}
\author{Gilles Nogues}
\author{Jo\"{e}l Cibert}
\affiliation{\NEELAddress}
\email[]{joel.cibert@neel.cnrs.fr}
\author{Kuntheak Kheng}
\affiliation{\IRIGAddress}
\author{David Ferrand}
\affiliation{\NEELAddress}

\date{\today}

\begin{abstract}
Quantum dots acting as single photon emitters in the blue-green range are fabricated and characterized at cryogenic temperature. They consist in CdSe dots inserted in (Zn,Mg)Se nanowires with a thick shell. Photoluminescence spectra, decay curves and autocorrelation functions were measured under nonresonant continuous-wave and pulsed excitation. An analytical approach is applied simultaneously to the decay curves and correlation functions. It allows a quantitative description of how these two quantities are affected by the exciton rise due to biexciton feeding, the bright exciton decay, the effect of the dark exciton, and the re-excitation between two laser pulses. Linewidths at our limit of resolution (200 $\mu$eV) are recorded. The reported correlation counts vary from a full control by re-excitation from traps, to a small contribution of re-excitation by mobile carriers or other QDs, as low as 5\%.
\end{abstract}

\maketitle

%\tableofcontents

\section{\label{sec:Intro}Introduction}

Semiconductor quantum dots (QD) are actively contemplated as single photon emitters for quantum communications \cite{Somaschi2016,Arakawa2020}. In this context, the main focus is put onto III-V dots grown by the Stranski-Krastanov method and emitting in the IR range, particularly in the so-called telecom window \cite{Senellart2021}. In spite of impressive achievements, a drawback of this system is the cryogenic temperatures which are needed to make the QDs  operate as efficient, pure single-photon emitters. CdSe QDs embedded in ZnSe appear as complementary since (1) they operate as single photon emitters \cite{Sebald2002,Tribu2008,Rakhlin2018,Rakhlin2021} up to room temperature \cite{Bounouar2012,Fedorych2012}, and (2) they emit in the blue-green range, a wavelength range with a specific interest for underwater or water-air communication \cite{Hufnagel2020,Li2019,Hu2019,Zhao2019}. Thus, these QDs appear as promising single-photon emitters for underwater or water-air quantum key distribution.

An attractive configuration for bright single-photon emission is that of a QD embedded in a waveguide \cite{Claudon2010}, or even in a nanowire (NW) \cite{Dalacu2019} which can be inserted in a photonic circuit \cite{Dalacu2021}. The shell around the QD acts as a waveguide and it constitutes also an efficient environment akin to protect the QD from surrounding defects. This configuration reduces the influence of neighboring QDs which could be excited by the same laser pulse and modify the dynamics of the selected QD \cite{Laferriere2021}. This is a clear advantage with respect to Stranski-Krastanov QDs.

In the case of CdSe QD in NWs, several configurations have been tested, including a QD with no shell  \cite{Bounouar2012}, or a QD with a shell added post-growth \cite{Jeannin2017}. In addition to the demonstration of room-temperature operation, several studies have explored the CdSe exciton dynamics \cite{Sallen2009}, the spectral diffusion within the electrostatically broadened zero-phonon line \cite{Sallen2010}, and the re-excitation by traps \cite{Aichele2004}. Here we address the most promising configuration, that of a self-assembled tapered shell grown together with the QD in a unique run of molecular beam epitaxy (MBE).

The first test of a single-photon emitter is the Hanbury Brown and Twiss experiment, \emph{i. e.}, the measure of the correlation function $g^{(2)}(t)$, either under continuous-wave (CW) or pulsed excitation. Both characterize the single-photon character, and its deviation due to overlap with multi-excitons or uncorrelated background. However, pulsed excitation is needed to characterize the ability to provide on-demand single-photon operation. We show here that the quantitative analysis of a combined measure of decay and correlation under pulsed excitation allows us to fully characterize the different components of the excitonic cascade and the associated signal such as re-excitation from traps, mobile carriers or adjacent QDs \cite{Aichele2004,Santori2004,Mnaymneh2019,Laferriere2020}, as well as spectral overlaps of the constituents of the radiative cascade \cite{Dalacu2020}. To this purpose, we extend the use of the rate equations and their analytical solutions generally used for CW-excitation data  \cite{Dalacu2020,Heindel2016,Moreau2001}, to the case of pulsed-excitation data which usually are analyzed through a probabilistic, numerical approach. We apply this approach to two CdSe QDs in thick tapered ZnSe NWs, a first one which exhibits a strong re-excitation by traps, and a second one which features very good on-demand single-photon emission properties.

The paper is organized as follows. Section~\ref{sec:Methods} describes the growth conditions and the characteristics of the selected samples, and describes the optical setup. Section~\ref{sec:CW} details the experimental results (spectra and correlations) under CW excitation. Section~\ref{sec:Pulsed} is devoted to the results (decay curves and correlations) obtained under pulsed excitation and to their quantitative analysis using the phenomenological analytical model developed in the Appendix. Section~\ref{sec:Discussion} summarizes and discusses the information acquired through this approach.

\section{\label{sec:Methods}Samples and experimental}

\subsection{\label{sec:Growth}Growth and samples}

%Figure Figure Figure Figure Figure Figure Figure Figure Figure Figure Figure Figure Figure Figure Figure Figure Figure
\begin{figure*}[]
\includegraphics [width=2\columnwidth]{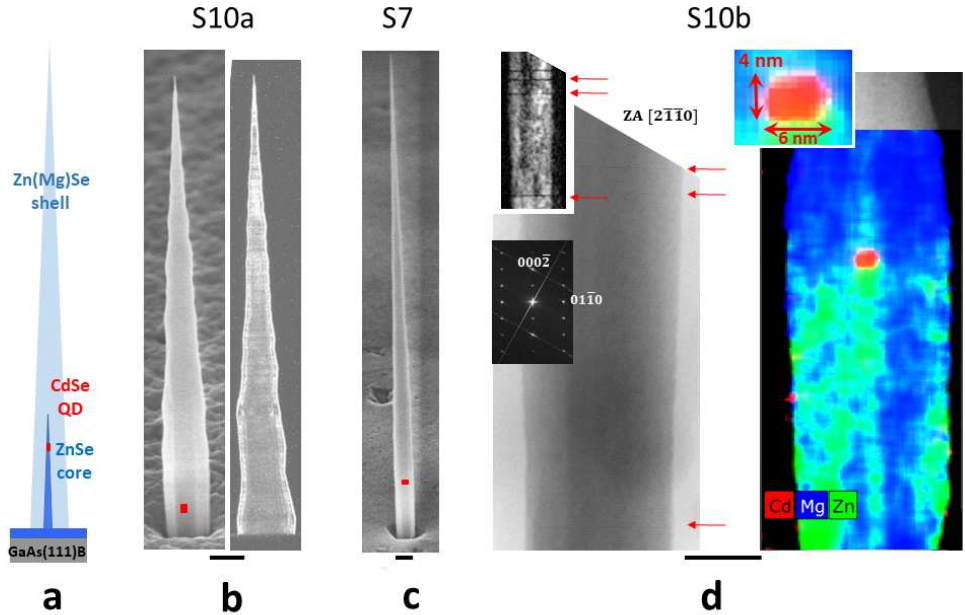}
\caption{\label{fig:fig1} (a) Schematic of the QD-core-shell structure; (b) Sample S10a with (left) the SEM image of a NW (at 65$^\circ$) and (right) the TEM darkfield image of another NW evidencing a large number of stacking faults; (c) Sample S7, SEM image (at 65$^\circ$); (d) Sample S10b; the left panel is the bright-field scanning-TEM image of a NW, showing (see the FFT in the inset) the wurtzite structure of the core of radius 8~nm and the 12~nm thick shell; The upper inset is the amplitude image of the $\{01\overline{1}0\}$ reflection, made with geometrical phase analysis \cite{Rouviere2005} on the same section of the NW, revealing more clearly the few stacking faults (red arrows). The right panel is the EDX map of another NW identifying the ZnSe core and the (Zn,Mg)Se shell, with a CdSe QD of 4~nm height and 3~nm radius (inset). Scale bars: 200~nm for (b) and (c), 20~nm for (d).}
\end{figure*}
%Figure Figure Figure Figure Figure Figure Figure Figure Figure Figure Figure Figure Figure Figure Figure Figure Figure

 A schematic of the whole structure is presented in Fig.\ref{fig:fig1}a. The growth conditions are detailed in Refs.~\cite{GosainPhD} and \cite{Gosain2022a}. Solid gold nanoparticles, with typical radius 3 to 5 nm, are formed on a ZnSe buffer layer, 8~nm thick, grown on a (111)B GaAs substrate \cite{Gosain2022a}. First, a ZnSe core is grown by molecular beam epitaxy at 350$^\circ$C. The nanowire radius at its tip is slightly smaller, by 0.5 nm, than the radius of the gold nanoparticle. The radial growth rate is very small so that the core is only weakly tapered \cite{Gosain2022a}. Then, the CdSe QD is inserted close to the top of the ZnSe core and finally a tapered ZnSe or (Zn,Mg)Se shell is grown at 300$^\circ$C or 320$^\circ$C.

We present the results of time-resolved spectroscopy obtained on two samples, S10a and S7. They differ by the duration of the Cd flux, 10~s and 7~s respectively (and hence by the QD height), and by some parameters of the shell, see Table~\ref{tab:table1}. Electron microscopy images are shown in Fig.\ref{fig:fig1}b and c, respectively. Scanning electron microscopy (SEM) imaging was performed using the secondary electrons in a Zeiss Ultra 55 (field emission gun) microscope, operated at 5–15 kV with typical beam currents in the 0.1–2.5 nA range. The sample tilt was 65$^\circ$. Transmission electron microscopy (TEM) sample preparation was performed by mechanical/wet dispersion on a holey carbon grid. TEM was performed on a CM300 working at 300 kV. Decreasing the growth temperature of the shell from 320$^\circ$C (sample S7 in Fig.\ref{fig:fig1}c) to 300$^\circ$C (sample S10a in Fig.\ref{fig:fig1}) slightly increases the radial growth rate but induces the formation of structural defects.

Another sample, S10b, was grown with a thin (Zn,Mg)Se shell in order to determine the profile of the CdSe QD using the quantitative analysis of Energy Dispersive X-ray Spectroscopy (EDX) described in Ref.~\cite{Rueda2016}. Scanning TEM and EDX was performed on a probe corrected Themis working at 200~kV. The result is shown in Fig.\ref{fig:fig1}d: It confirms the presence of a ZnSe core containing a (Cd,Zn)Se insertion with more than 50\% Cd, both with a radius $\sim3$~nm which reasonably matches the radius of the nanoparticle determined in Ref.~\cite{Gosain2022a}. The QD height is 4~nm. We expect the same size for sample S10a, which contains the same CdSe insertion (10 s of growth), and a 3~nm height for sample S7 with a 7~s insertion.

The overall crystal structure is wurtzite, with a good epitaxial relationship between the core and the shell (Fig.~\ref{fig:fig1}d). The possibility of a zinc-blende structure in the QD and around has been demonstrated in CdSe-ZnSe nanowires with a thin diameter \cite{DenHertog2011}. It gives rise to an increased thickness of the shell at the level of the QD. Although we did not observe such a thickening in the present S10a and S7 samples, the structure of a small QD with a thick shell is difficult to assess.

\begin{table}[b]
\caption{List of samples, with measured base radius (in nm), shell growth temperature (in $^\circ$C), Cd-cell opening time (in s) for the growth of the QD, and measured or expected QD height (in nm).}\label{tab:table1}%
\begin{ruledtabular}
\begin{tabular}{llllll}
Sample &Shell & Base  & Growth &Cd &QD \\
 & content&  radius & Temp. &duration & height\\
 &  & (nm) &($^\circ$C)&(s)&(nm)\\
\colrule
S10a &(Zn,Mg)Se& 140 & 300 &10 &4 expected\\
S10b&(Zn,Mg)Se& 30 & 320 &10 &4 measured\\
S7 &ZnSe& 90 & 320 &7& 3 expected\\
\end{tabular}
\end{ruledtabular}
\end{table}

\subsection{\label{sec:Experimental}Spectroscopy set-up}

The as-grown samples were mounted on a cold-finger cryostat cooled down to 5-6 K, and a single nanowire was excited along its axis, and its photoluminescence (PL) detected along the same axis in a confocal setup. CW excitation was provided by a laser diode emitting at 405 nm (a photon energy 3.06 eV, larger than the ZnSe bandgap), focused through a long working distance microscope objective of numerical aperture NA=0.55. This results in a laser spot of about 1 $\mu$m in diameter on the sample, small enough to excite a single nanowire in our samples with a low nanowire density. The 0.46 m spectrometer was equipped with a 1800 grooves/mm grating providing a resolution around 0.7 meV with a slit width 0.2 mm, and slightly less than 0.2 meV with the smallest slitwidth (0.05 mm). The detection was ensured by a charge-coupled device (CCD).

Pulsed excitation was provided by a near infra-red picosecond Titanium-Sapphire laser (pulse duration smaller than 2 ps and repetition time $T_0$=13.1 ns), doubled in frequency to 440 nm (2.82 eV) using a frequency doubling $\beta$-BaB$_2$O$_4$ crystal. Spectra shown here were recorded with the same setup as for CW excitation. For time-resolved data, detection was provided by single-photon avalanche photodiodes (APD) id100-50 from id-Quantique, mounted on the side exit of the spectrometer. The same setup was used to measure the decay of luminescence, or the correlation functions in the Hanbury Brown and Twiss (HBT) configuration, using a Becker and Hickl time-correlated single photon counting (TCSPC) module. The spectrometer was equipped with the 1800 grooves/mm grating or a 600 grooves/mm grating, providing a passband from 0.7 to 3 meV full width at half maximum (FWHM).  The time resolution of the setup is limited by the response time of the fast APD's, which is reasonably well described by a Gaussian function of 60~ps FWHM \cite{Becker2005}, with a few-ns long diffusion tail of intensity less than 1\% of the fast pulse \cite{Becker2005}. For the two arms of the Hanbury Brown and Twiss setup, the response is thus given by a Gaussian function $\frac{1}{\sigma\sqrt{2\pi}}\exp (-\frac{t^2}{2\sigma^2})$ with standard deviation $\sigma=$40~ps.

\section{\label{sec:CW}Spectroscopy and dynamics under CW excitation}
\subsection{\label{sec:Spectra} Spectra}

%Figure Figure Figure Figure Figure Figure Figure Figure Figure Figure Figure Figure Figure Figure Figure Figure Figure
\begin{figure*}
\includegraphics [width=2\columnwidth]{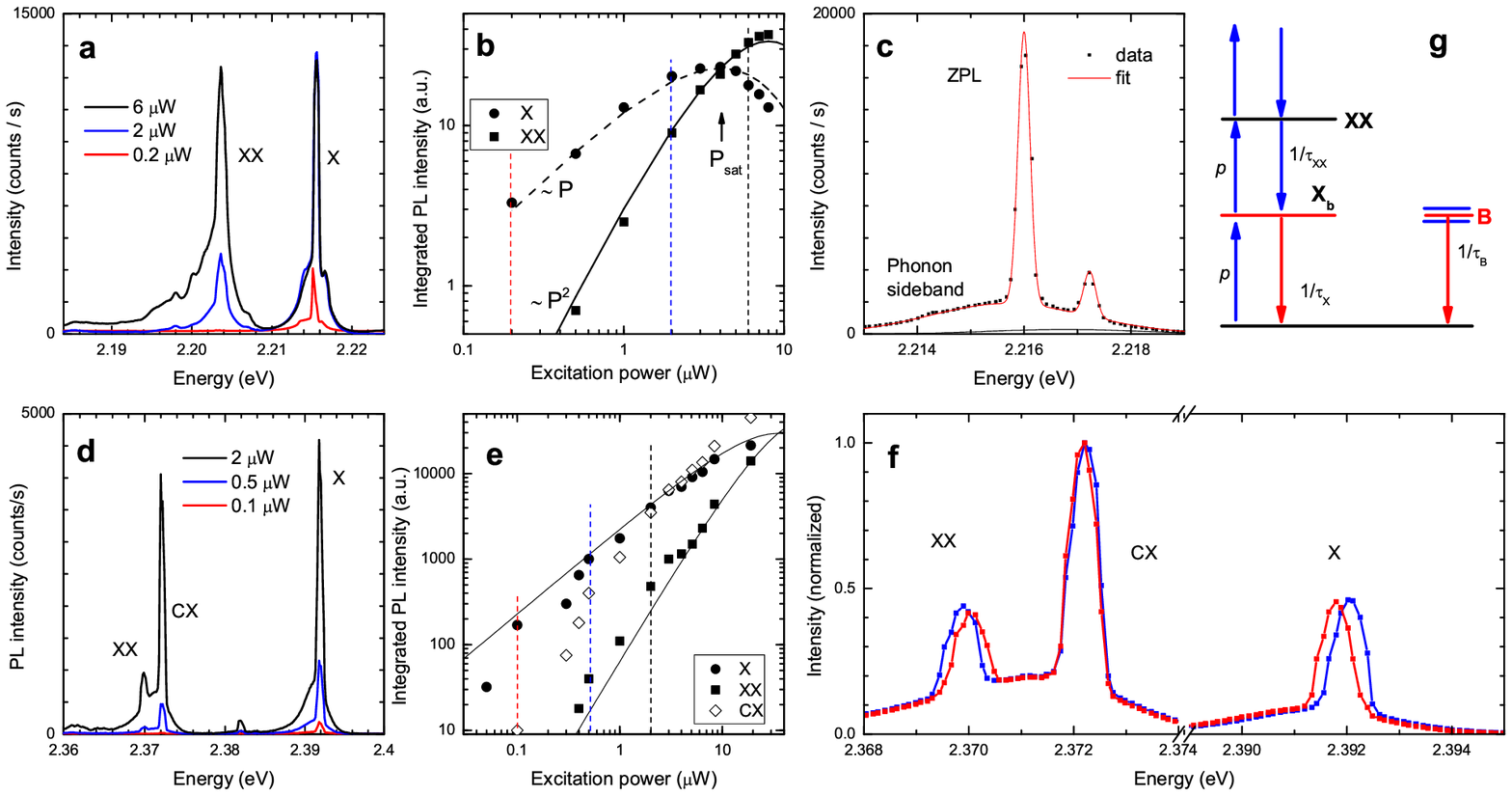}
\caption{\label{fig:fig2} Photoluminescence of sample S10a (top) and sample S7 (bottom), with CW excitation at 3.06 eV, detection CCD and grating 1800 grooves/mm. (a) to (c) sample S10a: (a) PL spectra for increasing values of the excitation power, as indicated, slits 0.2 mm; (b) PL intensity of the two main lines in (a), as a function of excitation power; the solid lines show the fit with Eq.~\ref{eq:eq1}; (c) PL spectrum at 5 $\mu$W excitation power and entrance slit 0.05 mm; In the fit (solid line), a Gaussian line of FWHM 260 $\mu$eV (the zero-phonon line, ZPL, at 2.216 eV) is associated to another Gaussian line of FWHM 3.3 meV (phonon sideband), with 40\% of the area in the ZPL. (d) to (f) Sample S7: (d) PL spectra for increasing values of the excitation power, as indicated, grating 1800 grooves/mm, slits 0.2 mm (e) PL intensity of the three main lines with vertical lines identifying the three spectra in (d), as a function of excitation power; (f) zoomed views of the three main lines for two orthogonal linear polarizations, excitation power 10 $\mu$W, grating 1800 grooves/mm, slits 0.2 mm, with the intensity normalized for the line at 2.372 eV. (g) Scheme of the different levels involved in our analysis of data under CW excitation. The spectrometer is tuned to the luminescence of the bright exciton $X_b$ (red arrow); Eq.~\ref{eq:eq1} involves the bright exciton $X_b$ and biexciton XX, and higher order excitons in a simple description. A part of the background light $B$ is recorded (second red arrow). The probability of excitation per unit time is described by $p$.}
\end{figure*}
%Figure Figure Figure Figure Figure Figure Figure Figure Figure Figure Figure Figure Figure Figure Figure Figure Figure

%Figure Figure Figure Figure Figure Figure Figure Figure Figure Figure Figure Figure Figure Figure Figure Figure Figure
\begin{figure*}
\includegraphics [width=2\columnwidth]{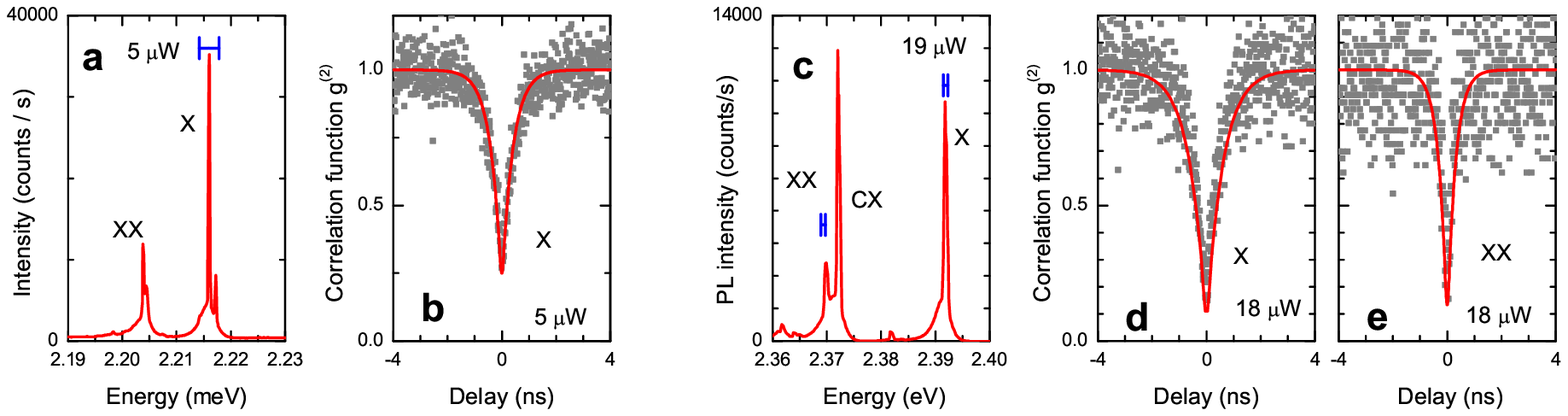}
\caption{\label{fig:fig3} Normalized photoluminescence autocorrelation $g^{(2)}(t)$. (a and b) Sample S10a with (a) spectrum, grating 1800 grooves/mm, slits 0.2 mm and CCD detection, the blue segment indicates the spectrometer position and passband used in the autocorrelation data in (b); (b) X autocorrelation, grating 600 grooves/mm, slits 0.3 mm and APD detection. (c to e) Sample S7, (c) spectrum, grating 1800 grooves/mm, slits 0.2 mm and CCD detection; the blue segments indicate the spectrometer position and passband used in the autocorrelation data; (d) X autocorrelation and (e) XX autocorrelation with grating 1800 grooves/mm, slits 0.2 mm and APD detection. The red curves are drawn using Eq.~\ref{g2CW}, with the adjustable parameters $T_{CW}=$0.4~ns and $\frac{B}{S}=0.10$ in (b), $T_{CW}=$0.6~ns and $\frac{B}{S}=0.02$ in (d) and $T_{CW}=$0.3~ns and $\frac{B}{S}=0.02$ in (e).}
\end{figure*}
%Figure Figure Figure Figure Figure Figure Figure Figure Figure Figure Figure Figure Figure Figure Figure Figure Figure

Figures~\ref{fig:fig2}a and d show three PL spectra recorded on a NW from sample S10a (top) and sample S7 (bottom), respectively, for different values of the excitation power. Each of the main lines comprises essentially a rather narrow component on top of a broader one. From the dependence on temperature \cite{GosainPhD}, we ascribe the broad component to the Stokes and anti-Stokes acoustic-phonon sidebands, and the narrow component to the zero-phonon line, as already observed for CdTe-ZnTe QDs \cite{Besombes2001} and CdSe-ZnSe ones \cite{Sebald2002}. Figures~\ref{fig:fig2}c shows a spectrum recorded with a better resolution. This high resolution spectrum confirms the presence of an additional, weak narrow line. The intensity and position of such lines, with respect to the main lines, varies from NW to NW. Their origin, parasitic QD in the same nanowire, or in a neighboring one, or in the 2D layer in between, or any other emitting center, is unknown. Note that a line such as that in Fig.~\ref{fig:fig2}c will remain out of the spectrometer passband in pulsed experiments with the 1800 grooves/mm grating, but not with the 600 grooves/mm grating (see the blue segment in Fig.~\ref{fig:fig3}a).

The attribution of the two main lines (and their sidebands) to the neutral exciton (X) and biexciton (XX) is deduced from the dependence of their intensity on the excitation power, linear and quadratic, respectively, see Fig.~\ref{fig:fig2}b and e. The whole set of data was fitted using Eq.~\ref{eq:eq1},
\begin{eqnarray}
I_X(P)&=&I_{sat} \frac{P}{P_{sat}} \exp (-\frac{P}{P_{sat}}) ,
\nonumber \\
I_{XX}(P)&=&I_{sat} \left(\frac{P}{P_{sat}}\right)^2 \exp (-\frac{P}{P_{sat}})
\label{eq:eq1}.
\end{eqnarray}

This expression results from the simple model described in Ref.~\onlinecite{Moreau2001} and schematized in the left part of Fig.~\ref{fig:fig2}g. This model assumes that the probability of decay of the $n$-exciton is proportional to $n$, for instance, the lifetime of the biexciton $\tau_{XX}$ is half that of bright exciton $\tau_X$; It neglects the effect of the dark exciton (we will show below that it plays a minor role at low temperature) and the more complex structure of the multi-excitons of higher order than the biexciton (which play a role only at excitation power much larger than $P_{sat}$). The two fitting parameters, $I_{sat}$ and $P_{sat}$, essentially depend on the experimental setup, but $P=P_{sat}$ means that the decay rate $\frac{1}{\tau_X}$ and the pumping rate $p$ of the bright exciton are equal.

The power dependence is not totally sufficient to decide which line is the biexciton line in sample S7, since two lines feature a superlinear power dependence. The final attribution is done by comparing the spectra for two orthogonal linear polarizations, Fig.~\ref{fig:fig2}f. A fine structure splitting, $\sim~200~\mu$eV, is observed with opposite signs on the two extreme lines (therefore attributed to X and XX), and not on the central line at 2.372 eV (therefore attributed to a charged exciton). A similar distribution in energy of the X, CX and XX lines, with a fine structure present, was observed in self-assembled CdSe QDs \cite{Patton2003} and excitons localized at CdSe-ZnSe local fluctuations \cite{Kummel1998}. An intermediate narrow line is visible on S7 at 2.382~eV in Fig.~\ref{fig:fig2}d (see also Fig.~\ref{fig:fig5}a below); this line was previously attributed to the negatively charged exciton \cite{Jeannin2017, Sallen2009}. Depending on the sample but also on the excitation conditions, we thus observe the two charged excitons with opposite signs.

The splitting between the X and XX lines (the so-called binding energy $B_{XX}$ of the biexciton) is larger in sample S7 (22 meV) than in sample S10a (15 meV). Similar values were found in other NWs from these two samples, and other samples.

We also observe additional lines for higher values of the excitation power. They are attributed to multi-excitons of index larger than 2.

A notable characteristics of the exciton lines is their linewidth, FWHM down to the 200~$\mu$eV range, and close to the limit of resolution of the setup in all spectra of Fig.~\ref{fig:fig2}.

\subsection{\label{sec:correlation} Autocorrelations}

Figure~\ref{fig:fig3} displays the normalized autocorrelation functions $g^{(2)}(t)$ recorded with CW excitation, from the same NWs as in Fig.~\ref{fig:fig2}, measured over the blue windows shown in the spectra recorded with the same excitation power (Figs.~\ref{fig:fig3}a and c). The excitation power was slightly below $P_{sat}$ as defined in Eq.~\ref{eq:eq1}.

All feature a simple anti-bunching behaviour, characterized \cite{Brouri2000} by a Laplace distribution \cite{Geraci2017}:
\begin{equation}
g^{(2)}(t)=1-\left[1-g^{(2)}(0)\right] \exp \left(-\frac{|t|}{T_{CW}}\right),
\end{equation}
where $T_{CW}$ is the time constant resulting from the dynamics of the exciton or biexciton, described \cite{Moreau2001} as a two-level system with an excitation rate $p$ and decay rate $\frac{1}{\tau}$, \emph{i.e.}, $\frac{1}{T_{CW}}=p+\frac{1}{\tau}$. In the ideal single-photon emitter, $g^{(2)}(0)=0$.

This response must be convoluted with the response function of the Hanbury Brown and Twiss setup. The result of the convolution of the Laplace distribution by a Gaussian function of standard deviation $\sigma$ is the normal-Laplace distribution, $GL(t)$ \cite{Geraci2017}, which is obtained by a straightforward calculation as
\begin{eqnarray}\label{GL}
&&GL(t))=\left[G(t)+G(-t)\right]\nonumber\\
&&G(t)=\frac{1}{2}\exp\left(\frac{\sigma^2}{2T_{CW}^2}+\frac{t}{T_{CW}}\right)\left[1-\erf \frac{1}{\sqrt{2}}\left(\frac{\sigma}{T_{CW}}+\frac{t}{\sigma}\right) \right],\nonumber\\
 \end{eqnarray}
where $\erf(z)=\frac{2}{\sqrt{\pi}}\int_0^z \exp (-t^2) dt$ is the error function. The measured autocorrelation function is thus
\begin{equation}\label{g2CW}
\tilde{g}^{(2)}(t)=1-\left[1-g^{(2)}(0)\right] \left[G(t)+G(-t)\right]
\end{equation}
To first order in $\frac{\sigma}{T_{CW}}$, $G(0)\simeq \frac{1}{2}-\sqrt{\frac{1}{2\pi}}\frac{\sigma}{T_{CW}}$ so that if $g^{(2)}(0)<<1$, $\tilde{g}^{(2)}(0)=g^{(2)}(0)+\sqrt{\frac{2}{\pi}} \frac{\sigma}{T_{CW}}$. The measured minimum is shifted upward by $\sqrt{\frac{2}{\pi}} \frac{\sigma}{T_{CW}}$.

A first contribution to non-vanishing of the zero-time correlation is the contribution of the background signal, which leads to replace $\left[1-g^{(2)}(0)\right]$ in Eq.~\ref{g2CW} by $\left[1-g^{(2)}(0)\right]\left(\frac{S}{S+B}\right)^2$ \cite{Brouri2000}, where $S$ is the signal from the single-photon emitter and $B$ the uncorrelated background. Note that this expression is obtained without reference to a specific model for the single-photon emitter. To first order in $\frac{B}{S}$, the minimum of $g^{(2)}(t)$ is shifted upward by $\frac{2B}{S}$.

The solid lines in Fig.~\ref{fig:fig3}b, d and e assume ideal single photon emitters, hence
\begin{equation}\label{g2CW}
\tilde{g}^{(2)}(t)=1-\left(\frac{S}{S+B}\right)^2 \left[G(t)+G(-t)\right].
\end{equation}
The values of the fitting parameters, $\frac{B}{S}$ and $T_{CW}$, are discussed in Section~\ref{sec:Discussion2}.

\section{\label{sec:Pulsed}Dynamics under pulsed excitation}

\subsection{\label{sec:pulsed results}Experimental results}

%Figure Figure Figure Figure Figure Figure Figure Figure Figure Figure Figure Figure Figure Figure Figure Figure Figure
\begin{figure*}
\includegraphics [width=2\columnwidth]{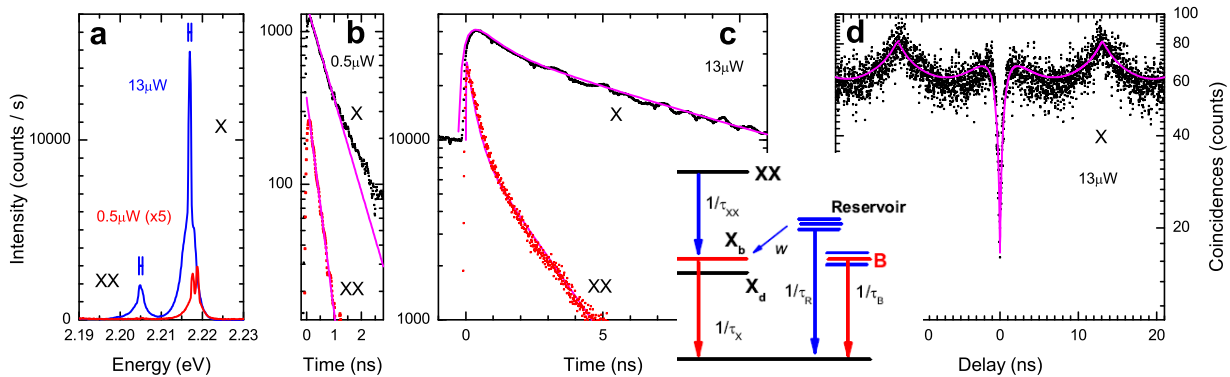}
\caption{\label{fig:fig4} Sample S10a under pulsed excitation, (a) spectra with 1800 grooves/mm, slits 0.2 mm, CCD detection, and (b to d) time-resolved data with grating 1800 grooves/mm, slits 0.2 mm and APD detection; (a) spectra at two different values of the excitation power, 0.5 $\mu$W (red curve) and 13 $\mu$W (blue); the blue segments indicate the spectrometer position and passband used in the time-resolved data; (b) decay for the exciton X (black symbols) and biexciton XX (red), with excitation power 0.5 $\mu$W, after subtraction of a constant baseline; the solid lines display exponential decay curves with characteristic times 0.7 and 0.3 ns, respectively (c) decay for X (black symbols) and XX (red), with excitation power 13 $\mu$W; the solid line shows the fit as described in text (d) exciton autocorrelation at 13 $\mu$W; the solid line shows the fit with the same parameters as in (c). The insed displays a schematic diagram of the different levels involved in our analysis of the pulsed-excitation data. The spectrometer is tuned to the luminescence of the bright exciton $X_b$; Eq.~\ref{eq:eq1} involves the biexciton XX and the bright exciton $X_b$; the analysis considers also the dark exciton $X_d$, background light $B$ (part of it being recorded, as schematized by the red arrow), and a reservoir $R$ with a probability per unit time $w$ of repopulating the QD if this one is empty.}
\end{figure*}
%Figure Figure Figure Figure Figure Figure Figure Figure Figure Figure Figure Figure Figure Figure Figure Figure Figure

\begin{figure*}
\includegraphics [width=2\columnwidth]{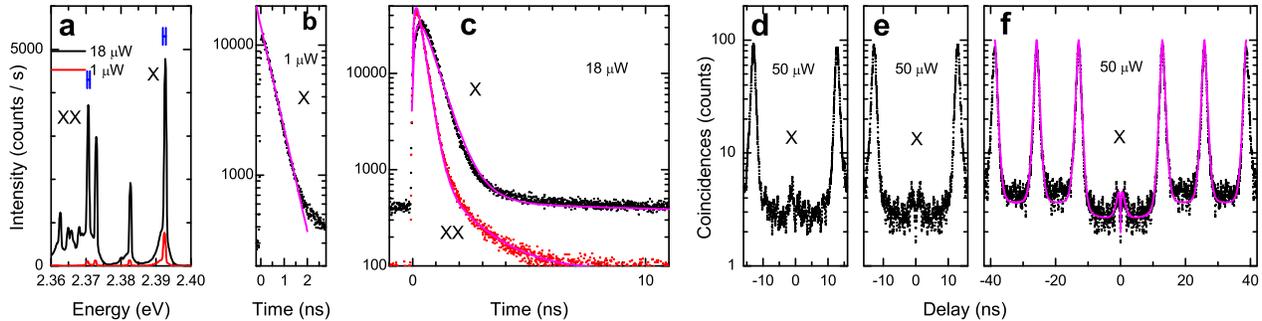}
\caption{\label{fig:fig5} Sample S7 under pulsed excitation, grating 1800 grooves/mm, slits 0.2 mm (a) spectra at two different values of the excitation power, 1 $\mu$W (red curve) and 18 $\mu$W (black) both with CCD; the blue segments indicate the spectrometer position and passband used in the time-resolved data; (b) decay for X (black symbols) with excitation power 1 $\mu$W, after subtraction of a constant baseline; the solid line displays an exponential decay with characteristic time 0.5 ns; (c) decay for X (black symbols) and XX (red), with excitation power 18 $\mu$W; the solid lines show the fit as described in text; (d) exciton autocorrelation at 50 $\mu$W; (e) same as (d) after symmetrization; (f) same as (e) with the fit (solid line) using the same parameters as in (c).}
\end{figure*}

Photoluminescence spectra under pulsed excitation are given in Fig.~\ref{fig:fig4}a for sample S10a and Fig.~\ref{fig:fig5}a for sample S7, for two different values of the excitation power. The exciton and biexciton lines, already identified with CW excitation, exhibit also the characteristic, linear or quadratic, increase with the pulsed excitation power as shown in Fig. 4-28 of Ref. \cite{GosainPhD} for another NW).

Figure~\ref{fig:fig4}b shows the evolution in time of the exciton and biexciton intensity after a low-power excitation pulse and under such conditions that the fine structure splitting is not resolved. A constant baseline has been subtracted. Both X and XX feature an immediate rise (faster than our time resolution of 60 ps), followed by an exponential decay with characteristic times 0.3 ns (XX) and 0.7 ns (X). A slow component is weak but visible in the decay of the exciton.

At high excitation power the exciton exhibits a slow rise, and the intensity of the slow component dramatically increases (Fig.~\ref{fig:fig4}c) so that the signal remains high even at the end of the arrival of the following pulse at $t=T_0$ (=13.1~ns). The whole decay is reasonably well described by a sum of three exponential functions, to be detailed below. We may note that the XX signal follows a similar trend, with a slow component which however is not so intense and not so slow as for the exciton signal.

Finally, the autocorrelation of the exciton $C(t)$ is quite singular (Fig.~\ref{fig:fig4}d), with a wide Laplace distribution at each non-zero delay $nT_0$, and another one around zero-delay with a narrow dip at its center. Similar shapes have been evidenced previously in III-V QDs \cite{Santori2004,Mnaymneh2019,Laferriere2020} and in CdSe QDs \cite{Aichele2004} and attributed to re-excitation effects. The fit will be described in the next subsection.

%Figure Figure Figure Figure Figure Figure Figure Figure Figure Figure Figure Figure Figure Figure Figure Figure Figure
%Figure Figure Figure Figure Figure Figure Figure Figure Figure Figure Figure Figure Figure Figure Figure Figure Figure

The same set of data is shown for sample S7 in Fig.~\ref{fig:fig5}. At low excitation power (Fig.~\ref{fig:fig5}b), the exciton signal exhibits a fast rise and a sub-ns decay as the main component, with characteristic time 0.6 ns, and a slower component more than a decade below. The slow component is clearly visible at higher excitation power (Fig.~\ref{fig:fig5}c) but even there, its intensity remains much weaker than the intensity of the fast component: this makes a clear contrast with the results for sample S10a. The  correlation function (Fig.~\ref{fig:fig5}d) is dominated by narrow Laplace-like peaks at each non-zero $nT_0$. It is necessary to use a log scale plot to obtain some information on the contributions around zero delay. The raw data (Fig.~\ref{fig:fig5}d) are not totally even with respect to time. This is not expected if the two arms of the Hanbury Brown and Twiss setup are totally equivalent, with the same spectral window. Here we must suspect that a radiative cascade takes place and is not recorded identically by the two arms, so that the signal is different at positive and negative delays. A simple example which could give rise to such an imbalance, is a stray signal due to the biexciton, or a multiexciton, on one arm. Another possible mechanism arises if the re-excitation involves a change in the charge distribution around the QD and a small shift of the emission line, with a different impact on the two arms. Fig.~\ref{fig:fig5}e displays the symmetrized signal, $\frac{1}{2}[C(t)+C(-t)]$: The shape suggests a re-excitation process. Fig.~\ref{fig:fig5}f shows the same symmetrized signal with the fit described below. Note also that the plateau between $t=0$ and $t=T_0$ is significantly lower than between $T_0$ and $2T_0$.

\subsection{\label{sec:pulsed model}Analytical model}
Immediately after the laser pulse at $t=0$, the QD may contain one or several electron-hole pairs. This excitation process takes place over a characteristic time in the sub-ps or ps range and is considered here to be infinitely short. The cascade which follows is a random process: there is a probability $\mathcal{P}_X(t)$ that the recombination of the last electron-hole pair takes place exactly at time $t$, and in the absence of non-radiative recombination, the signal $I_X(t)$ recorded at the single exciton wavelength is proportional to the average of $\mathcal{P}_X(t)$ over laser pulses. The ratio $I_X/\mathcal{P}_X$ incorporates all experimental characteristics (escape of photons from the sample, collection on the front lens and throughput of the optical set-up, efficiency of the detector and electronics, accumulation time). In order to calculate $I_X/\mathcal{P}_X$, appropriate factors should be introduced to take into account the possibility of non-radiative recombination. The knowledge of these factors is not required if, as done in the present approach, we restrict ourselves to the phenomenological description of $I_X$.

In the experimental conditions of non-resonant excitation for on-demand single-photon emission, the QD contains several electron-hole pairs after each laser pulse. As a result, we assume that the initial distribution is restored at each laser pulse at $nT_0$, so that the intensity $I_X(t)$ is periodic, with period $T_0$.

The intensity signal corresponding to the excitonic cascade in the QD can generally be adequately reproduced by a sum of three exponential functions, repeated after each pulse. In the interval $[0,T_0]$,
\begin{equation}\label{Eq_cascade}
 I_X(t)=\sum_{i=1}^3 A_i \exp(-\frac{t}{\tau_i}).
\end{equation}

The parameters entering Eq.~\ref{Eq_cascade} can be calculated using a master equation restricted to the exciton+biexciton system, \emph{i.e.}, the QD contains no electron-hole pair, one electron-hole pair (bright and dark exciton) or two electron-hole pairs, as displayed in the left part of the inset of Fig.~\ref{fig:fig4}. With respect to the CW case (Fig.~\ref{fig:fig2}g), we omit the multiexciton states (which could play a role at higher excitation power), keeping only X and XX, but we added the dark exciton (which may be revealed at long delay). And of course there is no laser excitation rate ($p=0$) since the laser pulse is over. The physical parameters involved in the dynamics are the bright exciton lifetime $\tau_X$ and the biexciton lifetime $\tau_{XX}$, to which one should add the dark exciton lifetime (due to decay through non radiative channels or through  mixing with the bright exciton), and spin-flip transitions between the dark and bright exciton levels \cite {Sallen2009,Santori2004,Moreau2001}. These parameters allow one to build the master equation which governs the population of the three excited levels. This master equation can be written in a vectorial form, $\frac{d\mathbf{n}}{dt}=\mathbf{M~\mathbf{n}}$, where the components of the vector $\mathbf{n}$ are the populations of the three excited states, and $\mathbf{M}$ is a $3\times3$ matrix. Therefore, whatever the number of processes involved, the full dynamics is described by three, and only three, characteristic time constants, which are obtained as the inverse of each of the three eigenvalues of $\mathbf{M}$.

This is the only piece of information that we need to keep in mind in the present study. However, we may note that with the mechanisms mentioned above, the master equation matrix splits into a contribution decribing the biexciton decay (and exciton rise, except at very low excitation) with $\tau_2=\tau_{XX}$, and a $2\times2$ matrix describing the bright-exciton / dark-exciton dynamics. The analytical solution is written explicitly for instance in Ref.~\cite{Dalgarno2005}. The bright exciton signal exhibits a decay with two exponential contributions. If the spin-flip process is slow compared to the exciton decay, there is a fast initial decay $A_1 \exp(-\frac{t}{\tau_1})$ with a time constant $\tau_1\simeq\tau_X$, and a slower one, $A_3 \exp(-\frac{t}{\tau_3})$ (the "dark exciton component"), which corresponds to the final decay of the bright+dark exciton, with a balanced population, governed by the spin-flip transition from the dark exciton to the bright exciton~\cite{Dalgarno2005}. In CdSe QDs, the dark exciton component has been observed to be weak at 4K and to become visible as the temperature increases \cite {Sallen2009,Patton2003}. In the opposite case of fast spin-flip, the initial decay is related to the spin-flip. In intermediate cases, and in more complex cases, the population of the three excited states is still described by three time constants $\tau_1$ to $\tau_3$, but their interpretation may differ from the simple ones.

Finally, the description requires at least two additional terms, so that the signal is
\begin{eqnarray}\label{Eq_Intensity}
  I(t)&=& I_X(t)+I_R(t)+I_B(t)\\
  I_R(t)&=&A_R \exp(-\frac{t}{\tau_R}),\nonumber\\
  I_B(t)&=&A_B \exp(-\frac{t}{\tau_B}).\nonumber
\end{eqnarray}

The second term, $I_R(t)$, describes the effect of re-excitation from entities which have been excited by the laser pulse and can re-populate the QD when the last photon of the cascade has been emitted. Re-excitation processes have been evidenced in CdSe QDs \cite{Aichele2004} as well as in III-V QDs \cite{Santori2004,Mnaymneh2019,Laferriere2020}. The source of re-excitation was attributed to excitation or charge traps  \cite{Aichele2004} or to decaying band-edge carrier population \cite{Mnaymneh2019,Laferriere2020}. We write $\tau_R$ the lifetime of this population, and $w$ the probability of re-excitation of the QD per unit time, see the inset in Fig.~\ref{fig:fig4}.

The last term, $A_B \exp(-\frac{t}{\tau_B})$ represents the background signal, \emph{i.e.}, non-correlated luminescence due to any parasitic photon source. It can be due to other QDs (then $\tau_B$ will be of the same order as $\tau_1$) or to any object emitting at the same wavelength (with any value of $\tau_B$), or straight light including laser light or a constant background.

In a Hanbury Brown and Twiss experiment, a first photon is detected at time $t_1$ and a second photon at time $t_2$, and the number of coincidences is recorded as a function of the delay time $t=t_2-t_1$. The first-photon signal is given by $I(t_1)$, and we can without loss of generality assume that $t_1\in [0,T_0]$ (this constitutes a definition of the pulse labelling). In order to calculate the coincidence count, we must distinguish the first period $[0,T_0]$, from the following ones $[nT_0,(n+1)T_0]$ with $n>0$.

If $t\in [T_0,2T_0]$, and with $t_1\in [0,T_0]$, then $t_2=t_1+t$ implies $t_2\in [T_0,3T_0]$. Within this interval, when writing the intensity at time $t_2$, we must distinguish the two sub-intervals $[T_0,2T_0]$ and $[2T_0,3T_0]$. Thus the coincidence count is proportional to
\begin{eqnarray} \label{Eq_Coinc1}
  C(t)&=& \int_0^{T_0}dt_1 I(t_1) \int_{T_0}^{2T_0}dt_2 I(t_2-T_0) \delta(t_2-t_1-t)  \nonumber\\
    &+&  \int_0^{T_0}dt_1 I(t_1)\int_{2T_0}^{3T_0}dt_2 I(t_2-2T_0) \delta(t_2-t_1-t)\nonumber\\
\end{eqnarray}
with $I(t)$ given by Eq.~\ref{Eq_Intensity}. As the QD content is reset to the same initial value at each pulse, Eq.~\ref{Eq_Coinc1}, shifted adequately, holds if $t\in [nT_0,(n+1)T_0]$.

If $t\in [0,T_0]$, the single-photon character has to be taken into account.  In this case, $t_2\in [0,2T_0]$ and we must distinguish between the two intervals, $[0,T_0]$ and $[T_0,2T_0]$.

If $t_2\in[T_0,2T_0]$, the QD has been re-excited by the laser pulse at $T_0$  between $t_1$ and $t_2$ and the first member of Eq.~\ref{Eq_Coinc1} applies.

If $t_2\in[0,T_0]$, then $I(t_2)$ is not given by Eq.~\ref{Eq_Intensity} since the different contributions, $I_X$, $I_B$ and $I_R$, behave differently after the emission of a photon at time $t_1$:
\begin{itemize}
  \item the QD is empty immediately after $t_1$, hence $I_X(t_2)=0$, there is no contribution until the next laser pulse;
  \item the non-correlated background is not altered by the emission of a photon from the QD, $I_B(t_2)=A_B \exp(-\frac{t_2}{\tau_B})$;
  \item the QD can be re-excited so that the $I_R$ contribution is restored, but with a risetime after the emission at time $t_1$.
\end{itemize}
  The effect of re-excitation is usually evaluated numerically through a stochastic approach \cite{Aichele2004,Santori2004,Mnaymneh2019,Laferriere2020}. We propose here an analytical description using the same approach as for CW correlations. The relevant part in the system described in the inset of Fig.~\ref{fig:fig4} consists in the empty QD, the single-exciton state of the QD, and an effective level representing the reservoir. Dynamics involves the exciton lifetime $\tau_1$, the lifetime $\tau_R$ of the reservoir population, and the probability of re-excitation probability per unit time $w$ by transfer from the reservoir. Immediately after time $t_1$, the QD is empty. If the population of the reservoir remains constant, the average QD population at time $t$ after the emission of the first photon is $w \tau [1-\exp(-\frac{t}{\tau})]$, with $\frac{1}{\tau}=w+\frac{1}{\tau_1}$. This is still reasonably valid if $w$ decreases slowly with the population of the reservoir, with the characteristic time $\tau_R$, so that the second-photon count at time $t_2$ is $I_R (t_2) [1-\exp(-\frac{t}{\tau})]$.

As a result, for $t\in [0,T_0]$,
\begin{eqnarray} \label{Eq_Coinc2}
  &C&(t)= \int_0^{T_0}dt_1 I(t_1)\int_{0}^{T_0}dt_2 I_B(t_2) \delta(t_2-t_1-t)\nonumber\\
  &+&  \int_0^{T_0}dt_1 I(t_1)\int_{0}^{T_0}dt_2 I_R(t_2)  \delta(t_2-t_1-t) [1-\exp(-\frac{t}{\tau})]\nonumber\\
  &+&\int_0^{T_0}dt_1 I(t_1) \int_{T_0}^{2T_0}dt_2 I(t_2-T_0) \delta(t_2-t_1-t)  \nonumber\\
\end{eqnarray}

The analytical calculation of Eq.~\ref{Eq_Coinc1} and \ref{Eq_Coinc2} with Eq.~\ref{Eq_Intensity} is given in the appendix. In short, if the decay signal is well-reproduced by a sum of exponential functions, the common expectation \cite{Miyazawa2016} that the coincidence count around each finite $nT_0$ is described by a sum of Laplace distributions is correct, but the characteristic time comes as a prefactor for each component. Hence slow components are enhanced. In addition, once again as expected but sometimes overlooked, special terms appear in the $[-T_0,T_0]$ segment, which ensures that the coincidence count vanishes at zero delay unless specific processes take place, such as multi-excitonic stray light or non-correlated emission.

If necessary, the Laplace distributions are replaced by normal-Laplace distributions in order to take into account the time resolution of the setup (as in Eq.~\ref{g2CW}). In the present case, this was found necessary only for the central dip of re-excitation.

For the two samples, a good fit of both the decay and coincidence counts is obtained with a single set of parameters. The relevant parameters, kept constant between decay and coincidence, are the decay time $\tau_1$, the rise time $\tau_2$ and relative amplitude $\frac{A_2}{A_1}$, the long component in the radiative cascade ("dark exciton") with time $\tau_3$ and relative amplitude $\frac{A_3}{A_1}$, the re-excitation with time $\tau_R$ and relative amplitude $\frac{A_R}{A_1}$, and the relative amplitude $\frac{A_B}{A_1}$ of background contribution assumed to be constant in time. All values are given in Table~\ref{tab:table2}. The vertical scales are left independent between the decay and coincidence plots. This also allows for changes of incident power.

\begin{table}[b]
\caption{Values of the five pairs of parameters used in Figs.~\ref{fig:fig4} and \ref{fig:fig5}. Three contributions $A_i \exp(-\frac{t}{\tau_i})$ are needed to describe the exciton luminescence in a system limited to three levels, bright exciton, dark exciton, biexciton. Re-excitation is described by adding another contribution, and background signal another one.}\label{tab:table2}%
\begin{ruledtabular}
\begin{tabular}{lcdr}
\textrm{Sample}&
\textrm{Contribution}&
\textrm{$A_i / A_1$}&
\textrm{$\tau_i$ (ns)}\\
\colrule
S10a & X fast decay ($i$=1)&  & 0.7\\
& X rise ($i$=2)& -1 & 0.3\\
& X slow decay ($i$=3)& 0 & \\
& X re-excitation ($R$)& 1.4 & 8\\
& Background ($B$)& 0.09 & $\infty$\\
\colrule
S7 & X fast decay($i$=1)& & 0.46\\
& X rise ($i$=2)& -1 & 0.23\\
& X slow decay ($i$=3)& 0.005&30 \\
& X re-excitation ($R$)& 0.05 & 0.8\\
& Background ($B$)& 0.001 & $\infty$\\
\end{tabular}
\end{ruledtabular}
\end{table}

\section{\label{sec:Discussion}Discussion}

\subsection{\label{sec:Discussion1}Spectroscopy of CdSe quantum dots}

Our identification of the biexciton line is based on the characteristic dependence on the excitation power, both in CW and pulsed excitation, but also on the energy shift with respect to the linear polarization, with a symmetric splitting observed on the neutral exciton and biexciton lines of several QDs.  This symmetric splitting is the signature of a fine structure splitting. Note that this observation is possible in the present axial configuration (made easy by the guiding effect of the tapered shell), while it was not in previous studies of CdSe QDs in nanowires with observation along a transverse direction. In the present samples its value varies from NW to NW and ranges from non-measurable (less than 100 $\mu$eV) to 400 $\mu$eV. The fine structure splitting was reported in self-assembled dots on (001) orientation \cite{Patton2003,Kulakovskii1999}. Here it is observed in spite of the (111) orientation, which suppresses a mechanism based on the crystal structure, and points to a deviation from a circular or hexagonal shape of the shell or the QD.

We observe a splitting between the neutral exciton and biexciton lines ranging from 15 meV in the thicker QDs (thickness 4 nm, nanowire S10a and other NWs from the same sample and other samples) to 22 meV in the shorter QDs (thickness 3 nm, sample S7a and other NWs from the same sample). This agrees with simple ideas on the role of confinement, with a larger binding energy in the smaller nanostructures. It could also points to a role of the piezoelectric field, likely to separate the electron and hole apart in the thicker QD, but the fact that these values are quite similar to previous results not only in QDs inserted in nanowires, but also in self-assembled QDs with the $<001>$ orientation \cite{Kulakovskii1999}, suggests that the effect of the piezoelectric field remains small.

We have attributed two other lines to the two types of charged excitons. These two lines are non polarized even if a fine structure splitting is present, and they exhibit a power dependence intermediate between X and XX. One, giving rise to the central line equidistant from the X and XX lines, appears to be associated with pre-existing carriers. This central line was sometimes attributed to negatively charged exciton \cite{Aichele2004,Seufert2003}. The other line is close to the XX line and exhibits a quadratic dependence on excitation power, which we ascribe to the photocreated nature of the associated charge. Such a charged exciton line close to XX was also occasionally observed \cite{Patton2003,Kummel1998}.

Although not a priority of this study, we observe linewidths much smaller than previously observed in CdSe QDs with a thin shell or a shell deposited post-growth, for instance 0.9 meV in Ref~\cite{Bounouar2012}, or even 2 meV, with a Gaussian profile, in QDs where spectral diffusion was studied by cross-correlations \cite{Sallen2010}. A thick shell is akin to protect the QD from the effect of surface traps. The present linewidths are even smaller than in self-assembled QDs, where the FWHM ranges from 1 meV \cite{Aichele2004} to 0.3 meV \cite{Ulrich2003}. More precise studies are needed to assess the actual linewidth of the present structures, including its temperature dependance which contains an information about the dephasing by phonons \cite{Denning2020}.

These narrow zero-phonon lines sit on top of broader bands associated to longitudinal acoustic phonon sidebands \cite{Sebald2002,Besombes2001}. The general behaviour of these sidebands is well-documented \cite{Luker2015,Denning2020}. The main parameter is the Huang-Rhys factor $S$, with the relative intensity of the zero-phon line given by $\exp (-S)$. The one-phonon sideband has a total intensity $S \exp (-S)$. It consists in two contributions. The anti-Stokes contribution on the high-energy side involves the absorption of a phonon, and is proportional to the population of acoustic phonons $n(E)$ where $E$ is the energy shift. The Stokes contribution on the low-energy side involves the creation of a phonon, with a probability proportional to $[n(E)+1]$. The evolution with temperature can be schematized as follows.

\begin{itemize}
  \item At 6K, the fit in Fig.~\ref{fig:fig2}c results in $S=0.7$. At this temperature, as $S<1$, the main contribution to the phonon sideband is the one-phonon contribution. One may note that the overlap between the exciton and biexciton lines, accompanied by such a phonon sideband, separated by 22 meV, is totally negligible, so that the full single exciton contribution may be recorded without any contribution from the biexciton. However, this will be detrimental to the observation of long-standing Rabi oscillations and to the indiscernibility of the emitted photons - a trade-off is unavoidable between brightness, which requires a spectral window as broad as possible, and indiscernibility which requires a spectral filtering.
  \item Decreasing the temperature below 6K will decrease the intensity of the phonon sidebands. The anti-Stokes component, proportional to $n(E)$, vanishes, while the Stokes component proportional to $[n(E)+1]$ decreases to a finite value which can be roughly estimated by subtracting the anti-Stokes component from the Stokes component at finite temperature. Applied to the fit of the sidebands in Fig.~\ref{fig:fig2}c, we expect a decrease of $S$ by a factor of 4, so that the zero-phonon line acquires 85\% of the total intensity. A further optimization can be looked for in the QD shape and size.
  \item  When increasing the temperature above 6K, the Huang-Rhys factor is expected to increase linearly with the temperature so that multiple-phonon contributions become significant, thus increasing the width of the phonon sideband up to 15 to 20 meV at room temperature \cite{Bounouar2012,GosainPhD}.
\end{itemize}

The NW system offers good opportunity for implementing pick-and-place techniques, with probably a possibility to include the NW in a resonant structure, but to our knowledge, this  was not explored up to now.

Further studies are needed on all these aspects which are beyond the scope of the present study.

\subsection{\label{sec:Discussion2}Three approaches to the dynamics}

We have described the results of three complementary approaches: the decay of photoluminescence after pulsed excitation, the correlation function under CW excitation, and the correlation function under pulsed excitation. They bring complementary pieces of information on the dynamics of the QD system.

The decay curves provide very precise information on the short-time processes. The fast-decay time of the exciton is in the sub-ns range, as expected for a CdSe QD \cite{Sebald2002,Bounouar2012,Sallen2009,Patton2003} . The present QDs are oriented along the polar $<111>$ axis and experience a mismatch strain from the ZnSe surrounding material, hence a piezoelectric field is built in and the electron and the hole of the excitonic pair are pushed away to opposite interfaces. However, detailed calculations in the parent system CdTe-ZnTe \cite{Moratis2021} show that for such flat QDs (radius 4 nm and height 4 nm) the shift is small, provided the valence and electron band offsets are not vanishingly small. Then the electron-hole  overlap is not dramatically reduced, and the probability of radiative recombination remains high, which offers good perspectives for fast communication close to GHz rates provided a good brightness is achieved. The fast-decay time of the biexciton is smaller than the fast-decay time of the exciton, with a ratio close to 2. This value is assumed in the simplest description of the biexciton, which was used to analyze the power dependence under CW excitation in section \ref{sec:Spectra}, assuming no non-radiative decay. Note that this result, $\tau_1 \simeq 2 \tau_{XX}$, suggests that the spin-flip process is slower than the bright-exciton decay, so that $\tau_1=\tau_X$ and $\tau_3$ corresponds to the spin-flip transition from the dark to the bright exciton \cite{Dalgarno2005}. An important piece of information in the context of quantum communication is that, in both samples, the risetime of the exciton signal ($\tau_2$) is equal to the short decay time of the biexciton signal, $\tau_2=\tau_{XX}$. In Figs.~\ref{fig:fig4}c and \ref{fig:fig5}c, the exciton rise amplitude is equal to its fast decay amplitude, $\frac{A_2}{A_1}=-1$ in the fit. The error bar on $\frac{|A_2|}{A_1}$ depends on the signal to noise ratio and on the time resolution of the setup, in the present conditions the ratio is definitely larger than 0.9. In the simple interpretation of the $I_1$ and $I_2$ components of the phenomenological description, $\frac{A_2}{A_1}=-1$  means that each laser pulse creates more than two electron-hole pairs in the QD. Hence, the on-demand condition is fulfilled. At lower P (Figs.~\ref{fig:fig4}b and \ref{fig:fig5}b), the absence of rise time, $\frac{A_2}{A_1}$ ratio close to zero, confirms that the pumping is too low to ensure on-demand operation. Finally, the origin of the slow signal, which is particularly visible at high excitation power, is difficult to ascertain from the simple decay curve: It may involve a dark exciton contribution, as well as a slow re-excitation or a background contribution.

The correlation functions $g^{(2)}(t)$ under CW excitation bring additional pieces of information. The characteristic times $T_{CW}$ are of the same order as $\tau_1$ from decay curves, or slightly smaller. This agrees with the fact that the correlation functions were recorded under such conditions that the XX signal is quite visible but smaller than the X signal, so that the pumping rate $p$ is smaller than $\frac{1}{\tau_1}$. Fitting with the normal-Laplace distribution shows that a large part of the apparent $g^{(2)}(0)$ value is due to the finite time resolution of our setup. To first order, the contribution is given by $\sqrt{\frac{2}{\pi}} \frac{\sigma}{T_{CW}}=0.08$, compared to $\tilde{g}^{(2)}(0)=0.25$ for the exciton in sample S10a (Fig.~\ref{fig:fig3}b), $\sqrt{\frac{2}{\pi}} \frac{\sigma}{T_{CW}}=0.05$ compared to $\tilde{g}^{(2)}(0)=0.09$ for the exciton in sample S7 (Fig.~\ref{fig:fig3}d), and  $\sqrt{\frac{2}{\pi}} \frac{\sigma}{T_{CW}}=0.10$ compared to $\tilde{g}^{(2)}(0)=0.13$ for the biexciton in sample S7 (Fig.~\ref{fig:fig3}e). From the spectra in Fig.~\ref{fig:fig3}, and the discussion on the phonon sideband, a biexciton contribution to $g^{(2)}(0)$ is likely to be negligible. We attribute what remains to a background contribution, which is quite small ($\frac{B}{S}\approx0.02$ in Figs.~\ref{fig:fig3}d and e) for Sample S7. It is larger ($\approx0.1$) in Fig.~\ref{fig:fig3}b for sample S10a, due at least in part to the broader detection window (with even a contribution from the additional line at 2.217 eV).

The correlation function under pulsed excitation carries a complementary information. Unlike the correlation under CW excitation, it can be measured under conditions relevant for on-demand single-photon emission. With respect to the simple decay curve, it is less accurate in the determination of the fast components (for instance, the biexciton feeding), but it contains the decisive information needed on the slow contributions and allows us to disentangle the dark exciton contribution, slow re-excitation and background signal through their characteristic shapes around zero delay.

To be more precise, the presence of the characteristic time $\tau_i$ as a prefactor of the Laplace distributions which constitute the coincidence peaks, makes them less sensitive to the existence of a risetime. Although the exciton rise is extremely visible in the initial part of the decay signal, the coincidence peak is merely broadened and the presence of the biexciton feeding could be simply overlooked. To take a simple example, let us assume that the signal contains only the exciton decay and the biexciton feeding, \emph{i.e.}, only $A_1$ and $A_2$ are non-zero, and consider the two extreme cases $A_2=0$ (no bi-exciton feeding) and $A_2=-A_1$ (complete biexciton feeding). In the first case, the decay signal is $I(t)=A_1 \exp(-\frac{t}{\tau_1})$ and the right-wing part of the Laplace distribution around $T_0$ is $C(t)=\frac{1}{2} A_1^2 \tau_1 \exp(-\frac{t-T_0}{\tau_1})$: these two profiles are similar. In the second case, with a typical $\tau_2=\frac{1}{2} \tau_1$, the decay signal is $I(t)=A_1 [\exp(-\frac{t}{\tau_1})-\exp(-\frac{2t}{\tau_1})] $ and the right-wing part of the Laplace distribution is $C(t)=\frac{1}{6} A_1^2 \tau_1 [\exp(-\frac{t-T_0}{\tau_1})-\frac{1}{2}\exp(-\frac{2(t-T_0)}{\tau_1})]$. While $I(t)$ features a clear signature of the biexciton feeding (it starts from zero and increases before decreasing with characteristic time $\tau_1$), $C(t)$ simply exhibits a flat shape, with a significantly reduced maximal value, and it is easily misinterpreted in the absence of the information from the decay curve.

The presence of the prefactor has of course the opposite effect of enhancing the slow component of the excitonic cascade. As a result, it significantly contributes to the plateaus between coincidence peaks at non-zero delays. The plateau around zero delay disappears. This happens not only because the Laplace peak at $t=0$ disappears, as sometimes assumed \cite{Miyazawa2016}, but also because the specific contributions are such that the total signal vanishes at $t=0$. The peculiar shape caused by this vanishing of the excitonic cascade contribution at zero-delay is exemplified in Fig.~\ref{fig:fig6}b for the dark exciton contribution, and in Fig.~\ref{fig:fig6}c for the re-excitation. The shapes of these two contributions are notably different, and provide us with a precise tool to identify each of them.

Our two samples are good examples of such an identification. A long component within the excitonic cascade signal is needed for the fit of S7 (Fig.~\ref{fig:fig5}. In the present data at low temperature, the contribution is small ($\frac{A_3}{A_1}$ is a fraction of percent, and we cannot exclude an artefact from a long tail of the APD response \cite{Becker2005}). The fit probably gives only the order of magnitude, but it cannot be assumed to vanish. The fit requires also the presence of some re-excitation, with $\frac{A_R}{A_1}$ of a few percent. The characteristic time, $\tau_R$ of the order of 1~ns, is essentially measured on the correlation signal but it also contributes to the decay and ignoring its contribution would affect our determination of $\tau_1$.

In sample S10a, the correlation signal around zero delay is dominated by the re-excitation process (Fig.~\ref{fig:fig4}). A possible dark-exciton contribution is totally masked at this temperature but could re-appear at slightly higher temperature, where this channel is expected to acquire significant values \cite{Sallen2009}.

Again, as in CW, we did not need to introduce a biexciton contribution , which should appear as correlation peak at zero-delay, and is expected to be of the order of $\frac{I_{XX}}{I_X}$ where $I_{XX}$ is the leak of biexciton signal at the exciton energy \cite{Nair2011}. This is too small to be detected due to the large binding energy from the spectra in Figs.~\ref{fig:fig2}a and c, but again should contribute at higher temperature as the linewidth increases.

Our analytical approach benefits from the simplifying assumption of the reset of the QD content to an initial value at each pulse. It uses only five parameter pairs. A straightforward extension is the possibility of a distribution of characteristic times for the re-excitation, as well as for the background signal. Although its implementation is straightforward, it is difficult to test in a simple way. The model is also easily extended in order to include the possibility of stray photons from the excitonic cascade, for instance biexciton or multiexciton photons detected within the window around the single exciton \cite{Nair2011}. They do not contribute in the present case.

The hypothesis of a steady state is reasonable for the re-excitation and the background signals, and also for the excitonic cascade in the present case of a strong excitation. A direct consequence is that the correlation peaks are of equal intensity. This assumption has to be questioned if the excitation is weak. Then, starting with an empty QD and assuming no re-excitation, the population at the end of each pulse is expected to increase from pulse to pulse towards the steady state population. In this case, we have to use the steady state values only for the decay signal and for the correlation contribution at time $t_1$ (the start photon). Immediately after $t_1$ the QD is empty and the population at time $t_2$ will build up at each laser pulse to eventually reach the steady-state value. The intensity of the correlation peaks follows that increase, which induces a modulation of their intensity, opposite to that which results from blinking \cite{Miyazawa2016,Santori2001,Dusanowski2017}.

Re-excitation was invoked previously in CdSe QDs \cite{Aichele2004}. Coming back to our two samples, the main difference between them is the presence of a large density ($\frac{A_R}{A_1}\simeq1$) of deep ($\tau_R=$8 ns) traps as a source of re-excitation in S10a: The X-signal is mostly fed by traps, so that the integrated spectrum displays a X-line stronger than XX. By contrast, the excitonic cascade dominates for S7, and there is only a trace of re-excitation , with a low value of $\frac{A_R}{A_1}=0.05$ and a contribution of a few \% to the coincidence count. In addition, the value of the time constant, $\tau_R=$1 ns, suggests free carriers or at most shallow traps, or neighbouring QDs. Note that the deep traps of S10a are not related to the structural defects due to the low-temperature growth since traps are also observed (not shown) in an intermediate sample with 10~s QD and shell as in S7, grown at 320$^\circ$C with no Mg. In both cases the characteristic time of the re-excitation cycle, $\tau=(w+\frac{1}{\tau_1})^{-1}$, see Eq.~\ref{Eq_Coinc2}, was kept equal to the exciton lifetime $\tau_1$.

Re-excitation is expected to strongly depend on the sample properties, on the temperature, and the excitation conditions. In all cases, the present approach is a promising tool. Finally, re-excitation can be minimized by resonant excitation \cite{Somaschi2016,Miyazawa2016} or even better by two-photon excitation of the biexciton \cite{Schweickert2018}.

The present approach is profitably applied to other examples from the literature, such as InAs-InP structures. It reproduces satisfactorily the presence of a significant re-excitation in the Fig.~4 of Ref.~\cite{Laferriere2021}; the characteristic time $\tau$ is in the sub-ns range, much shorter than the exciton lifetime $\tau_1=1.5$~ns. In Fig.~10 of Refs.~\cite{Dalacu2021}, no re-excitation is visible but the use of a logarithmic scale and the simultaneous measure of the decay curve would allow a better evaluation of the presence of a slow component (dark-exciton type) at the center of the correlation plot. A good example is Fig. 3 of Ref.~\cite{Miyazawa2016}, where the logarithmic scale allows one to notice a deviation of the experimental signal from the proposed fit to a formula obtained by simply cancelling the Laplace peak at zero delay.

\section{\label{sec:Conclusion} Conclusion}

We propose an analytical expression of the coincidence count associated to the single-photon emission of a QD under pulsed pumping, taking into account the excitonic cascade and re-excitation. The calculation includes an explicit determination of the Laplace-distributions forming each coincidence peak at non-zero delay, built on the same exponential functions which describe the photoluminescence decay signal, and a proper treatment of the coincidence count around zero delay. Exploiting the link between the coincidence curve and the decay signal reveals complementary aspects of the dynamics of the excitons in a QD in relationship with on-demand single-photon emission. The time-dependence of the photoluminescence signal brings a precise information on the fast processes, such as the exciton decay and its feeding through the bi-exciton. It helps in determining the conditions for achieving the on-demand regime of single-photon emission. In a complementary way, the coincidence count is highly sensitive to slow processes such as the influence of the dark exciton, to re-excitation processes, and of course to the single-photon character of the emission. This complementary approach should benefit to the analysis of the single-photon character of the emission of various systems, including III-V QDs emitting in the telecom band. As an example, it is applied here to two CdSe QDs inserted in tapered ZnSe nanowires, so that their photoluminescence can be excited and recorded in a confocal configuration along the nanowire axis. The role of re-excitation is clearly evidenced in the $g^{(2)}(t)$ curve around $t=0$ and is quantified thanks to the phenomenological analytical approach. Flat QDs exhibit a large splitting, 22 meV, between the exciton line and the biexciton line, thus increasing the purity of the single-photon emitter.

As they emit in the blue-green range, these QDs in nawires appear as promising for underwater or air-to-sea quantum key distribution. This wavelength realizes a compromise between transmission in air in spite of a Rayleigh scattering larger than in the infra-red, and transmission in seawater (over tens of meters) in spite of the presence of turbulence \cite{Hufnagel2020, Hu2019}. The role of a highly fluctuating sea-air interface was also described in \cite{Li2019}. An operation of the QDs at room temperature is feasible but calls for a further optimization of the structure and of the excitation / detection conditions, and a confirmation of the brightness. The CW power dependance at low temperature was fitted without dark exciton contribution and without assuming any additional non-radiative channel. This is also supported by the pulsed excitation data at low temperature, including the values of the decay times of the exciton and biexciton and the weak, slow "dark exciton" contribution. The structure features other favourable aspects, such as the absence of a wetting layer which could capture the QD population when rising the temperature, and the presence of a well-matched tapered shell to collect a large part of the emitted photons along the NW axis. Solutions exist to redirect the backward emission. However, the precise measure of the brightness remains to be done.

In addition, the linewidths observed at low temperature are smaller than previously achieved in CdSe QDs in nanowires and in self-assembled CdSe QDs. The limits for the emission of undiscernible photons thus appear as less stringent as previously assumed and need to be further explored.

\begin{acknowledgments}
SRG acknowledges the European Union Horizon 2020 research and innovation programme under the Marie Sk{\l}odowska-Curie grant agreement No 754303 (GREnoble QUantum Engineering). The authors thank the CEA "Programme Exploratoire Bottom-Up" for financial support. We benefitted
from the access to the nano characterization platform (PFNC) in CEA Minatec Grenoble.
\end{acknowledgments}

\appendix

%Figure Figure Figure Figure Figure Figure Figure Figure Figure Figure Figure Figure Figure Figure Figure Figure Figure
\begin{figure}[b]
\includegraphics [width=1\columnwidth]{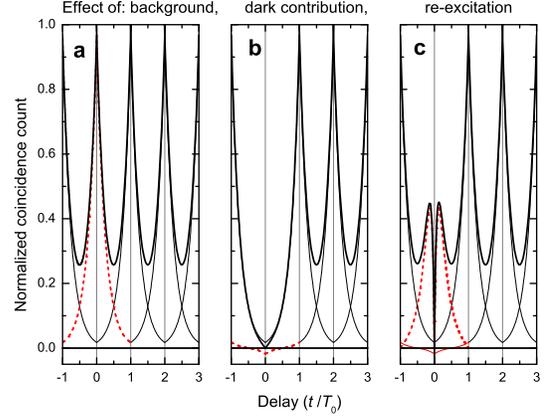}
\caption{\label{fig:fig6} (a) $C_{1B}(t)+C_{B1}(t)$ contribution, normalized, with $\tau_1=\frac{T_0}{4}$ and $\tau_B=\infty$ (thick solid curve); the red dashed curve is the Laplace distribution peaked at $t=0$, the thin solid curves are the other Laplace distributions; (b) $C_{11}(t)$ with $\tau_1=\frac{T_0}{4}$; the Laplace distribution centered at $t=0$ is replaced by the contribution shown by the red dashed curve; (c) $C_{RR}(t)$ with $\tau_R=\frac{T_0}{4}$ and $\tau=\frac{T_0}{10}$; the additional contribution is shown by the red dashed curve.}
\end{figure}
%Figure Figure Figure Figure Figure Figure Figure Figure Figure Figure Figure Figure Figure Figure Figure Figure Figure

\section{Details of the calculation}

The starting point is the probability of the difference between two variables $t_1$ and $t_2$, each having an exponential distribution $f_i(t_i)=\frac{1}{\tau_i}\exp(-t_i/\tau_i)$ for $t_i>0$ and 0 for $t_i<0$. This joint density is described by the asymmetric Laplace distribution,
\begin{eqnarray}\label{Laplace}
f_L(t)&=&\int_0^\infty f_1(t_1) dt_1 \int_0^\infty f_2(t_2) dt_2 \delta(t_2-t_1-t) \nonumber\\
&=& \frac{1}{\tau_1+\tau_2}\exp(t/\tau_1), t<0\nonumber\\
&=& \frac{1}{\tau_1+\tau_2}\exp(-t/\tau_2), t>0.
\end{eqnarray}
The Laplace distribution is peaked at $t=0$, with a characteristic constant $\tau_1$ on the negative side and $\tau_2$ on the positive side.

The calculation of Eq.~\ref{Eq_Coinc1} and \ref{Eq_Coinc2} involves essentially two modifications with respect to this simple case: (1) The exponential distributions are repeated at each multiple of $T_0$ (which induces Laplace-type peaks at each $nT_0$ and renormalization factors), and (2) a special treatment must be applied to the first interval $t_2\in[0,T_0]$, which implies a special treatment for $t\in[-T_0,T_0]$.

As $C(t)$ is an even function of $t$ we consider positive values of $t$. A straightforward calculation leads to:
\begin{equation}\label{Result1}
C(t)=\sum_{i,j}C_{ij}(t)
\end{equation}
with the $C_{ij}(t)$ as follows.
\begin{widetext}
In an interval $t\in[nT_0,(n+1)T_0]$ with $n>0$ an integer,
\begin{eqnarray}
C_{ij}(t)+C_{ji}(t)&=&A_iA_j\frac{\tau_i\tau_j}{\tau_i+\tau_j}
\left[\exp\left(-\frac{t-nT_0}{\tau_i}\right)+\exp\left(\frac{t-(n+1)T_0}{\tau_i}\right)\right]\left[1-\exp\left(-\frac{T_0}{\tau_j}\right)\right]\nonumber\\
&+&A_iA_j\frac{\tau_i\tau_j}{\tau_i+\tau_j}\left[\exp\left(-\frac{t-nT_0}{\tau_j}\right)+\exp\left(\frac{t-(n+1)T_0}{\tau_j}\right)\right]\left[1-\exp\left(-\frac{T_0}{\tau_i}\right)\right]
\end{eqnarray}
for $i,j=1$ to 3, $R$ and $B$, including the diagonal terms $C_{ii}$.
We recognize in the first line the right-hand side of Laplace distribution with characteristic time $\tau_i$ peaked at $nT_0$, followed by the left-hand side of the same distribution peaked at $(n+1)T_0$. Note also that the time constants enter the contribution to $C(t)$ as prefactors, so that the long-lived contributions will be enhanced in the coincidence count.

The same expression holds in the first interval, $t\in[0,T_0]$, for coincidences involving the (non-correlated) background contribution:
\begin{eqnarray}
C_{iB}(t)+C_{Bi}(t)&=&A_iA_B\frac{\tau_i\tau_B}{\tau_i+\tau_B} \left[\exp\left(-\frac{t}{\tau_i}\right)+\exp\left(\frac{t-T_0}{\tau_i}\right)\right]\left[1-\exp\left(-\frac{T_0}{\tau_B}\right)\right]\nonumber\\
                   &+&A_iA_B\frac{\tau_i\tau_B}{\tau_i+\tau_B} \left[\exp\left(-\frac{t}{\tau_B}\right)+\exp\left(\frac{t-T_0}{\tau_B}\right)\right]\left[1-\exp\left(-\frac{T_0}{\tau_i}\right)\right]
\end{eqnarray}
for $i=1$ to 3, including the diagonal term $C_{BB}$.

The coincidence count $C(t)$ thus contains Laplace distributions peaked at each $nT_0$. As an example, Fig.~\ref{fig:fig6}a shows, for $(C_{1B}+C_{B1}=$, the Laplace distribution centered at $t=0$ (red dashed line), specific to the background contributions, and the other ones at non-zero $nT_0$ (thin black line), relevant for all contributions. The coincidence count is the sum of all these contributions (thick black line). It keeps finite values between two peaks if the time constants are not infinitely small with respect to $T_0$.

The expression in $[0,T_0]$ is different for the coincidences involving the QD exciton cascade and the re-excitation. The Laplace distribution centered at $t=0$ disappears, and additional terms ensure that the contributions to $C(t)$ identically vanish at $t=0$.

For the excitonic cascade in $[0,T_0]$:
\begin{equation}\label{}
C_{ij}(t)+C_{ji}(t)=A_iA_j\frac{\tau_i\tau_j}{\tau_i+\tau_j}\left[\exp\left(\frac{t-T_0}{\tau_i}\right)-\exp\left(-\frac{t}{\tau_i}-\frac{T_0}{\tau_j}\right)+\exp\left(\frac{t-T_0}{\tau_j}\right)-\exp\left(-\frac{t}{\tau_j}-\frac{T_0}{\tau_i}\right)\right] \end{equation}
for $i,j=1$ to 3, including the diagonal terms $C_{ii}$. This can be rewritten as:
\begin{eqnarray}\label{}
&C&_{ij}(t)+C_{ji}(t)=A_iA_j\frac{\tau_i\tau_j}{\tau_i+\tau_j}\left\{\exp\left(\frac{t-T_0}{\tau_i}\right)\left[1-\exp\left(-\frac{T_0}{\tau_j}\right)\right]+\exp\left(\frac{t-T_0}{\tau_j}\right)\left[1-\exp\left(-\frac{T_0}{\tau_i}\right)\right]\right\}\nonumber\\
&+&A_iA_j\frac{\tau_i\tau_j}{\tau_i+\tau_j}2\left\{\sinh\left(\frac{t-T_0/2}{\tau_i}\right)\exp\left(-\frac{T_0}{2\tau_j}\right)+\sinh\left(\frac{t-T_0/2}{\tau_j}\right)\exp\left(-\frac{T_0}{2\tau_i}\right)\right\}\exp\left(-\frac{T_0}{2\tau_i}\right)\exp\left(-\frac{T_0}{2\tau_j}\right)
 \end{eqnarray}
which evidences in the first line the left-hand side of the Laplace distributions centered at $t=T_0$, and in the second line an additional contribution which ensures vanishing at $t=0$ (shown as a red dashed line in Fig.~\ref{fig:fig6}b for $C_{11}(t)$). This differs from the mere suppression of the central peak \cite{Miyazawa2016}.

Finally, the re-excitation contributions comprise another additional term:
\begin{eqnarray*}\label{}
&C&_{iR}(t)+C_{Ri}(t)=A_iA_R\frac{\tau_i\tau_R}{\tau_i+\tau_R} \left\{\exp\left(\frac{t-T_0}{\tau_i}\right)\left[1-\exp\left(-\frac{T_0}{\tau_R}\right)\right]+\exp\left(\frac{t-T_0}{\tau_R}\right)\left[1-\exp\left(-\frac{T_0}{\tau_i}\right)\right]\right\}\nonumber\\
&+&A_iA_R\frac{\tau_i\tau_R}{\tau_i+\tau_R} 2\left\{\sinh\left(\frac{t-T_0/2}{\tau_i}\right)\exp\left(-\frac{T_0}{2\tau_R}\right)+\sinh\left(\frac{t-T_0/2}{\tau_R}\right)\exp\left(-\frac{T_0}{2\tau_i}\right)\right\}\exp\left(-\frac{T_0}{2\tau_i}\right)\exp\left(-\frac{T_0}{2\tau_R}\right) \nonumber\\
&+&A_iA_R\frac{\tau_i\tau_R}{\tau_i+\tau_R}\left[1-\exp\left(-\frac{t}{\tau}\right)\right]\left[\exp\left(-\frac{t}{\tau_R}\right)-\exp\left(\frac{t-T_0}{\tau_i}-\frac{T_0}{\tau_R}\right)\right]
\end{eqnarray*}
for $i=1$ to 3, and including the diagonal terms $C_{RR}$. The additional term  (third line) is shown in red in Fig.~\ref{fig:fig6}c for $C_{RR}(t)$). It appears essentially as a Laplace distribution centered at $t=0$, with the contribution close to $t=0$ cut by the re-excitation factor with characteristic time $\tau$.

\end{widetext}

%------------------------------------------------------------------------------
%\section{References}

\begin{thebibliography} {}


\bibitem{Somaschi2016}
N. Somaschi, V. Giesz, L. De Santis, J. C. Loredo, M. P. Almeida, G. Hornecker, S. L. Portalupi, T. Grange, C. Ant\'{o}n, J. Demory, C. G\'{o}mez, I. Sagnes, N. D. Lanzillotti-Kimura, A. Lema\^{\i}tre, A. Auffeves, A. G. White, L. Lanco, and P. Senellart,
Near-optimal single-photon sources in the solid state,
Nature phot. \textbf{10}, 340 (2016).

\bibitem{Arakawa2020}
Yasuhiko Arakawa and Mark J. Holmes,
Progress in quantum-dot single photon sources for quantum information technologies: A broad spectrum overview,
Appl. Phys. Rev. \textbf{7}, 021309 (2020).

\bibitem{Senellart2021}
Pascale Senellart,
Semiconductor single-photon sources: progresses and applications,	
Photoniques \textbf{107}, 40 (2021).

\bibitem{Sebald2002}
K. Sebald, P. Michler, T. Passow and D. Hommel, G. Bacher, and A. Forchel,
Single-photon emission of CdSe quantum dots at temperatures up to 200K,
Appl. Phys. Lett. \textbf{81}, 2920 (2002).			

\bibitem{Tribu2008}
A. Tribu, G. Sallen, T. Aichele, R. Andr\'{e}, J.-Ph. Poizat, C. Bougerol, S. Tatarenko, and K. Kheng,		
A high-temperature single-photon source from nanowire quantum dots,
Nano Lett. \textbf{8}, 4326 (2008).

\bibitem{Rakhlin2018}
M. V. Rakhlin, K. G. Belyaev, S. V. Sorokin, I. V. Sedova, D. A. Kirilenko, A. M. Mozharov, I. S. Mukhin, M. M. Kulagina, Yu. M. Zadiranov, S. V. Ivanov, and A. A. Toropov,
Single-Photon Emitter at 80 K Based on a Dielectric Nanoantenna with a CdSe/ZnSe Quantum Dot,
JETP Lett. \textbf{108}, 201 (2018).

\bibitem{Rakhlin2021}
M. Rakhlin, S. Sorokin, D. Kazanov, I. Sedova, T. Shubina, S. Ivanov, V. Mikhailovskii, and A. Toropov,
Bright Single-Photon Emitters with a CdSe Quantum Dot and Multimode Tapered Nanoantenna for the Visible Spectral Range,
Nanomaterials \textbf{11}, 916 (2021).

\bibitem{Bounouar2012}
S. Bounouar, M. Elouneg-Jamroz, M. den Hertog, C. Morchutt, E. Bellet-Amalric, R. Andr\'{e}, C. Bougerol, Y. Genuist, J.-Ph. Poizat, S. Tatarenko, and K. Kheng,
Ultrafast Room Temperature Single-Photon Source from Nanowire-Quantum Dots,
Nano Lett. \textbf{12}, 2977 (2012).

\bibitem{Fedorych2012}
O. Fedorych, C. Kruse, A. Ruban, D. Hommel, G. Bacher, and T K\"{u}mmell,
Room temperature single photon emission from an epitaxially grown quantum dot,
Appl. Phys. Lett. \textbf{100}, 061114 (2012).

\bibitem{Hufnagel2020}
F. Hufnagel, A. Sit, F. Bouchard, Y. Zhang, D. England, K. Heshami, B. J. Sussman, and E. Karimi,
Investigation of underwater quantum channels in a 30 meter flume tank using structured photons,
New J. Phys. \textbf{22}, 093074 (2020).

\bibitem{Li2019}
Dong-Dong Li, Qi Shen, Wei Chen, Yang Li, Xuan Han, Kui-Xing Yang , Yu Xu, Jin Lin, Chao-Ze Wang, Hai-Lin Yong, Wei-Yue Liu, Yuan Cao, Juan Yin, Sheng-Kai Liao, and Ji-Gang Ren,
Proof-of-principle demonstration of quantum key distribution with seawater channel: towards space-to-underwater quantum communication,
Opt. Communic. \textbf{452}, 220 (2019).

\bibitem{Hu2019}
Cheng-Qiu Hu, Zeng-Quan Yan, Jun Gao, Zhi-Qiang Jiao, Zhan-Ming Li, Wei-Guan Shen, Yuan Chen, Ruo-Jing Ren, Lu-Feng Qiao, Ai-Lin Yang, Hao Tang, and Xian-Min Jin,
Transmission of photonic polarization states through 55-m water: towards air-to-sea quantum communication,
Photonics Res. \textbf{7}, A40 (2019).

\bibitem{Zhao2019}
Shicheng Zhao, Wendong Li, Yuan Shen, YongHe Yu, XinHong Han, Hao Zeng, Maoqi Cai, Tian Qian, Shuo Wang, Zhaoming Wang, Ya Xiao, and Yongjian Gu,
Appl. Optics 58, 3902 (2019).
Experimental investigation of quantum key distribution over a water channel

\bibitem{Claudon2010}
J. Claudon, J. Bleuse, N. S. Malik, M. Bazin, P. Jaffrennou, N. Gregersen, C. Sauvan, P. Lalanne, and J.-M. G\'{e}rard,
A highly efficient single-photon source based on a quantum dot in a photonic nanowire,
Nat. Photonics \textbf{4}, 174 (2010).

\bibitem{Dalacu2019}
D. Dalacu, P. J Poole, and R. L. Williams,
Nanowire-based sources of non-classical light,
Nanotechnology \textbf{30}, 232001 (2019).

\bibitem{Dalacu2021}
D. Dalacu, P. J. Poole, and Robin L. Williams,
Tailoring the Geometry of Bottom-Up Nanowires: Application to High Efficiency Single Photon Sources,
Nanomaterials \textbf{11}, 1201 (2021).

\bibitem{Laferriere2021}
P. Laferri\`{e}re, E. Yeung, I. Miron, D. B. Northeast, S. Haffouz, J. Lapointe, M. Korkusinski, P. J. Poole, R. L. Williams, and D. Dalacu,
Unity yield of deterministically positioned quantum dot single photon sources,
ArXiv 2110.08366	 https://doi.org/10.48550/arXiv.2110.08366

\bibitem{Jeannin2017}
M. Jeannin, T. Cremel, T. H\"{a}yrynen, N. Gregersen, E. Bellet-Amalric, G. Nogues, and K. Kheng,
Enhanced photon extraction from a nanowire quantum dot using a bottom-up photonic shell,
Phys. Rev. Applied \textbf{8}, 054022 (2017).

\bibitem{Sallen2009}
G. Sallen, A. Tribu, T. Aichele, R. Andr\'{e}, L. Besombes, C. Bougerol, S. Tatarenko, K. Kheng, and J.-Ph. Poizat,
Exciton dynamics of a single quantum dot embedded in a nanowire,
Phys. Rev B \textbf{80}, 085310 (2009).
					
\bibitem{Sallen2010}
G. Sallen, A. Tribu, T. Aichele, R. Andr\'{e}, L. Besombes, C. Bougerol, M. Richard, S. Tatarenko, K. Kheng, and J. Ph. Poizat,
Subnanosecond spectral diffusion measurement using photon correlation,
Nat. Photonics \textbf{4}, 696 (2010).

\bibitem{Aichele2004} 		
T. Aichele, V. Zwiller, and O. Benson,		
Visible single-photon generation from semiconductor quantum dots,
New J. Phys. \textbf{6}, 90 (2004).

\bibitem{Santori2004}  		
C. Santori, D. Fattal, J. Vuckovic, G. S. Solomon, and Y. Yamamoto,
Single-photon generation with InAs quantum dots			
New J. Phys. \textbf{6}, 89 (2004).

\bibitem{Mnaymneh2019}  		
K. Mnaymneh, D. Dalacu,  J. McKee, J. Lapointe, S. Haffouz, J. F. Weber, D. B. Northeast, P. J. Poole, G. C. Aers, and R. L. Williams,				
On‐Chip Integration of Single Photon Sources via Evanescent Coupling of Tapered Nanowires to SiN Waveguides,
Adv. Quantum Tech 1900021 (2019).

\bibitem{Laferriere2020}
P. Laferri\`{e}re, E. Yeung, L. Giner, S. Haffouz, J. Lapointe, G. C. Aers, P. J. Poole, R. L. Williams, and D. Dalacu,
Multiplexed Single-Photon Source Based on Multiple Quantum Dots Embedded within a Single Nanowire,
Nano Lett. \textbf{20}, 3688 (2020).

\bibitem{Dalacu2020}
D. Dalacu, D. B. Northeast, P. J. Poole, G. C. Aers, R L. Williams, K. A. Owen, and D. Oblak
Phys. Rev. B 102, 115401 (2020)	
Pump power control of photon statistics in a nanowire quantum dot	

\bibitem{Heindel2016}
T. Heindel, A. Thoma, M. von Helversen, M. Schmidt, A. Schlehahn, M. Gschrey, P. Schnauber, J.-H. Schulze, A. Strittmatter, J. Beyer, S. Rodt, A. Carmele, A. Knorr and S. Reitzenstein,
A bright triggered twin-photon source in the solid state,
Nat. Commun. \textbf{8}, 14870 (2017).	

\bibitem{Moreau2001}
E. Moreau, I. Robert, L. Manin, V. Thierry-Mieg, J. M. G\'{e}rard, and I. Abram,		
Quantum Cascade of Photons in Semiconductor Quantum Dots,
Phys. Rev. Lett. \textbf{87}, 183601 (2001).	

\bibitem{GosainPhD}
S. R. Gosain,
Room temperature single-photon source based on semiconductor quantum-dot nanowire for integrated photonics,
PhD Uni. Grenoble-Alpes (2021) https://hal.archives-ouvertes.fr/tel-03551997/

\bibitem{Gosain2022a}
S. R. Gosain, E. Bellet-Amalric, M. den Hertog, R. Andr\'{e}, and J. Cibert,
Nanotechnology https://doi.org/10.1088/1361-6528/ac5cfa
The onset of tapering in the early stage of growth of a nanowire

\bibitem{Rueda2016}
P. Rueda-Fonseca, E. Robin, E. Bellet-Amalric, M. Lopez-Haro, M. Den Hertog, Y. Genuist, R. Andr\'{e}, A. Artioli, S. Tatarenko, D. Ferrand, and J. Cibert,
Quantitative Reconstructions of 3D Chemical Nanostructures in Nanowires,
Nano Lett. \textbf{16}, 1637 (2016).

\bibitem{DenHertog2011}
M. Den Hertog, M. Elouneg-Jamroz, E. Bellet-Amalric, S. Bounouar, C. Bougerol, R. Andr\'{e}, Y. Genuist, J. Ph. Poizat, K. Kheng, and S. Tatarenko,
Insertion of CdSe quantum dots in ZnSe nanowires: Correlation of structural and chemical characterization with photoluminescence,
J. Appl. Phys. \textbf{110}, 034318 (2011).	

\bibitem{Becker2005}
TCSPC Performance of the id100-50 detector,
Becker and Hickl GmbH (2005), https://www.photonicsolutions.co.uk/upfiles/id100-50-becker.pdf

\bibitem{Rouviere2005}
J. L. Rouvi\`{e}re and E. Sarigiannidou,
Theoretical discussions on the geometrical phase analysis	,
Ultramicroscopy \textbf{106}, 1 (2005).	
	
\bibitem{Besombes2001}
L. Besombes, K. Kheng, L. Marsal, and H. Mariette,
Acoustic phonon broadening mechanism in single quantum dot emission,
Phys. Rev. B \textbf{63}, 155307, 2001.

\bibitem{Patton2003}
B. Patton, W. Langbein, and U. Woggon,
Trion, biexciton, and exciton dynamics in single self-assembled CdSe quantum dots,
Phys. Rev. B \textbf{68}, 125316 (2003).

\bibitem{Kummel1998}
T. K\"{u}mmell, R. Weigand, G. Bacher, A. Forchel, K. Leonardi, D. Hommel, and H. Selke,
Single zero-dimensional excitons in CdSe/ZnSe nanostructures,
Appl. Phys. Lett. \textbf{73}, 3105 (1998).

\bibitem{Brouri2000}
R. Brouri, A. Beveratos, J.-Ph. Poizat, and Ph. Grangier,
Photon antibunching in the fluorescence of individual color centers in diamond,
Opt. Lett. \textbf{25}, 1294 (2000).

\bibitem{Geraci2017}
M. Geraci,
Mixed effects models using the Normal and the Laplace distributions,
https://doi.org/10.48550/arXiv.1712.07216

\bibitem{Dalgarno2005}
P. A. Dalgarno, J. M. Smith, B. D. Gerardot, A. O. Govorov,  K. Karrai, P. M. Petroff, and R. J. Warburton,
Dark exciton decay dynamics of a semiconductor quantum dot	
Phys. Stat. Sol. (a) \textbf{202}, 2591 (2005).

\bibitem{Miyazawa2016}
T. Miyazawa, K. Takemoto, Y. Nambu, S. Miki, T. Yamashita, H. Terai, M. Fujiwara, M. Sasaki, Y. Sakuma, M. Takatsu, T. Yamamoto, and Y. Arakawa,
Single-photon emission at 1.5 $\mu$m from an InAs/InP quantum dot with highly suppressed multi-photon emission probabilities,
Appl. Phys. Lett. \textbf{109}, 132106 (2016).
	
\bibitem{Kulakovskii1999}
V. D. Kulakovskii, G. Bacher, R. Weigand, T. K\"{u}mmell, A. Forchel, E. Borovitskaya, K. Leonardi, and D. Hommel,
Fine Structure of Biexciton Emission in Symmetric and Asymmetric CdSe/ZnSe Single Quantum Dots,
Phys. Rev. Lett. \textbf{82}, 1780 (1999).

\bibitem{Seufert2003}
J. Seufert, M. Rambach, G. Bacher, and A. Forchel,
Single-electron charging of a self-assembled II–VI quantum dot,
Appl. Phys. Lett. \textbf{82}, 3946 (2003).

\bibitem{Ulrich2003}
S. M. Ulrich, S. Strauf, P. Michler, G. Bacher, and A. Forchel,
Triggered polarization-correlated photon pairs from a single CdSe quantum dot,
Appl. Phys. Lett. \textbf{83}, 1848 (2003).

\bibitem{Denning2020} E. V. Denning, J. Iles-Smith, N. Gregersen, and J. Mork,
Phonon effects in quantum dot single-photon sources,	
Optical Materials Express \textbf{10}, 222 (2020)	

\bibitem{Luker2015} S. Luker, T. Kuhn, and D E Reiter,
Direct optical state preparation of the dark exciton in a quantum dot,
Phys. Rev. B \textbf{92}, 201305(R) (2015)	

\bibitem{Moratis2021}
K. Moratis, J. Cibert, D. Ferrand, and Y.-M. Niquet,
Light-hole states in a strained quantum dot: numerical calculation and phenomenological models,
Phys. Rev. B \textbf{103}, 245304 (2021).

\bibitem{Nair2011}
G. Nair, J. Zhao, and M. G. Bawendi,
Biexciton Quantum Yield of Single Semiconductor Nanocrystals from Photon Statistics	
Nano Lett. \textbf{11}, 1136 (2011).

\bibitem{Santori2001}   C. Santori, M. Pelton, G. Solomon, Y. Dale, and Y. Yamamoto,
Triggered Single Photons from a Quantum Dot,
Phys. Rev. Lett. \textbf{86}, 1502 (2001).	

\bibitem{Dusanowski2017} 	{\L}. Dusanowski, P. Holewa, A. Mary\'{n}ski, A. Musia{\l}, T. Heuser, N. Srocka, D. Quandt, A. Strittmatter, S. Rodt, J. Misiewicz, S. Reitzenstein, and G. S\c{e}k
 Triggered high-purity telecom-wavelength single-photon generation from p-shell-driven InGaAs/GaAs quantum dot,
Optics Express \textbf{25}, 31122 (2017).

\bibitem{Schweickert2018}
L. Schweickert, K. D. J\"{o}ns, K. D. Zeuner, S. F. Covre da Silva, H. Huang, T. Lettner, M. Reind, J. Zichi, R. Trotta, A. Rastelli, and V. Zwiller,
On-demand generation of background-free single photons from a solid-state source,
Appl. Phys. Lett. \textbf{112}, 093106 (2018).

\end{thebibliography}

%------------------------------------------------------------------------------

\end{document}